\documentclass[a4paper,12pt]{article}
\pdfoutput=1
\usepackage[T1]{fontenc}
\usepackage[utf8]{inputenc}
\usepackage{natbib}
\usepackage[english,french]{babel}
\usepackage{graphicx}
\usepackage{hyperref}

\begin{document}
\bibliographystyle{apalike}
\setcitestyle{aysep={}}
\title{The Poincaré pear and Poincaré-Darwin
  fission theory in astrophysics, 1885--1901\\
  {\tiny \emph{Philosophia Scientiae} 27(3), 159--187; doi
  \href{https://dx.doi.org/10.4000/philosophiascientiae.4178}{10.4000/philosophiascientiae.4178}}}
\author{Scott A.~Walter}
\date{2023-09-14}
\maketitle

\begin{abstract}
  Henri Poincaré a découvert dans les années 1880 une figure
d'équilibre d'une masse fluide en rotation uniforme : la figure
piriforme, ou la poire. Il considérait que la poire était susceptible
de se diviser en deux parties inégales, et que son évolution pouvait
expliquer la genèse des étoiles binaires. Les études photométriques et
spectroscopiques des étoiles variables de cette période ont donné lieu
à la modélisation en étoiles binaires à éclipses, ce qui renforçait
l'interprétation réaliste des figures d'équilibre -- y compris la
poire -- dans le domaine cosmique. Cet article propose une analyse de
l'interprétation astrophysique de la poire de Poincaré et de la
théorie Poincaré--Darwin sur la genèse des étoiles binaires par fission,
en regard des recherches sur les étoiles variables réalisées entre
1885 et 1901.
\end{abstract}

\selectlanguage{english}
\begin{abstract}
  In the early 1880s, Henri Poincaré discovered a new equilibrium
  figure for uniformly-rotating fluid masses -- the pear, or piriform
  figure -- and speculated that in certain circumstances the pear
  splits into two unequal parts, and provides thereby a model for the
  origin of binary stars. The contemporary emergence of photometric
  and spectroscopic studies of variable stars fueled the first models
  of eclipsing binaries, and provided empirical support for a realist
  view of equilibrium figures -- including the pear -- in the cosmic
  realm. The paper reviews astrophysical interpretation of the Poincaré
  pear and the Poincaré--Darwin fission hypothesis with respect to
  research on variable stars from 1885 to 1901.
\end{abstract}

\section{Introduction: scientific knowledge, far from equilibrium}

In the foreword to his exhaustive monograph on variable stars,
published in 1924, the Vatican astronomer Father Johan Stein (S.J.)
observed that a straightforward theory of these objects was
``virtually unimaginable''.\footnote{``Denn soweit sich das ganze Feld
  überschauen läßt, sind die physikalischen Ursachen der
  Veränderlichkeit so verschieden und verwickelt, daß man versucht
  ist, eine Formulierung, die den Keplerschen Gesetzen an Einfachheit
  auch nur von weitem gleichkäme, für geradezu undenkbar zu halten.''
  \citet[V]{SteJ1924}.} Contemporary observers were agreed. Variable
stars had been observed for centuries, and the periodicity of their
maximum luminosity tabulated, but it was only with the discovery of
spectroscopic binary stars in 1889 that their orbital elements were
estimated, and their origins and evolution imagined. Most notably,
periodic luminosity variation was largely understood by astronomers to
stem from eclipsing binary stars. Just a few years before the first
discovery of a spectroscopic binary, Henri Poincaré suggested that
binary stars find their origin in the fission of a
member of a new series of equilibrium figures: the pear, or piriform
figure \citep{hp1885ama}.

Poincaré's effort to prove the stability of the pear, and those of
Aleksandr Mikhailovich Liapunov to prove the contrary, are well-known
in outline to scientists and
historians \citep{JarW1958}. In his
history of equilibrium figures, the astrophysicist Subrahmanyan
Chandrasekhar wrote memorably of the ``grand mental panorama'' created
by Poincaré's discovery of the pear and its role in cosmogony, which
was ``so intoxicating that those who followed Poincaré were not to
recover from its pursuit'' \citep[11]{ChaS1969}.

The panorama and
intoxication of the Poincaré pear are explored in this paper from an
epistemological standpoint that engages with two questions: (1) how
did the fission theory come to prominence in astrophysics and
cosmology, and (2) what was Poincaré's own engagement with
research on eclipsing binary stars?  These are the
questions I address in some detail for the period beginning with
Poincaré's discovery of the piriform figure in 1885, and ending in
1901, when Poincaré first publicly embraced the emergent theory of
eclipsing binaries.

General histories of astronomy typically underline how a
class of variable stars, the Cepheids, came to provide a reliable
distance gauge both in the Milky Way and beyond, in the
clusters and galaxies of our galactic neighborhood. A
distinction was not made between eclipsing binaries and Cepheid-type
variables until around 1913, when a pulsation model was introduced for
the latter. This was an important moment for the eclipse model, as it
no longer had to account for the light curves of Cepheid
variables. Even so, theorists continued to deploy variants of the
eclipse hypothesis to explain Cepheid light curves well into the
1920s.\footnote{General histories of astronomy with useful accounts of
  eclipsing binaries include \citet{PanA1961} and
  \citet{HerD1984}. Primary sources in astronomical photometry and
  spectroscopy are well-referenced by \citet{HeaJ1996,HeaJ2014}. For a
  brief, non-technical history of variable stars see \citet{HogH1984}.}

My account begins with a brief, non-technical summary of Poincaré's
claim to have discovered a figure of equilibrium, known as the pear,
or piriform figure, and its interpretation in astrophysics in the
1890s as a preliminary stage in the genesis of binary stars. The
second section takes up George William Myers' model of the variable
star $\beta$ Lyrae as an eclipsing binary, while the third and final
section recalls Charles André's presentation of Myers' model,
Poincaré's reviews of André's textbook, and a letter from Myers to
Poincaré.

\section{The Poincaré pear and the Poincaré-Darwin fission theory}

Isaac Newton assumed the Earth to be a rotating figure of equilibrium,
such that the universal force of gravitation counter-balanced the
forces of centrifugal acceleration of the spinning globe. Newton found
the consequent figure of the Earth, assuming homogeneity, to be an
oblate spheroid.\footnote{Newton assumed a homogeneous figure of
  radius $R$ rotating with angular velocity $\Omega$ and ellipticity
  $\epsilon$. Denoting the gravitational constant $G$ and the
  terrestrial mass $M$, Newton related ellipticity to the centripetal
  acceleration at the equator, $\Omega^2 R$, divided by the mean
  gravitational acceleration at the surface, $GM/R^2$, such that
  $\epsilon = \frac{5}{4}\frac{\Omega^2 R^3}{GM}$. For accounts of
  Newton's line of reasoning, see \citet{GuiN1989,GreJ1995}.}
Measurement of the variations of gravitational force with terrestrial
latitude, and of the apparent form of Jupiter by
G.-D.~Cassini and John Flamsteed tended to confirm Newton's model, while
the mathematical theory was developed by Colin Maclaurin and others in the
eighteenth century.

In the first half of the nineteenth century, Carl Gustav Jacob Jacobi showed that,
contrary to intuition, there are figures of equilibrium of rotating,
homogeneous fluid masses that are \emph{not} figures of revolution,
including ellipsoids of \emph{three} unequal axes. In the 1880s,
Poincaré took an interest in this mathematical problem, inspired in
part by the analysis of the stability of Saturn's rings by Sofia
Kovalevskaia, as he explained -- anonymously -- in the pages of the
Parisian journal \emph{Le Temps} in 1886. This was also the occasion
to announce his own discovery of a new series of equilibrium figures:
the apiodal, or piriform figure of equilibrium. His description appealed
to the imagination:
\begin{quote}
  Imagine a fluid mass contracting via cooling so slowly that it
  remains homogeneous, and its angular velocity is the same at every
  point. From a nearly spherical form it flattens more and more, while
  remaining an ellipsoid of revolution. Then the equator itself ceases
  to be circular, becoming elliptical; the fluid mass assumes the form
  of an ellipsoid with three unequal axes. Next, the median part of
  the ellipsoid thins out; one of the halves tends to elongate more
  and more, and the other half tends toward a spherical
  form. Everything leads us to believe that if the cooling were to
  continue, our fluid mass would divide in two distinct and unequal
  bodies.\footnote{``Imaginons une masse fluide, se contractant par
    refroidissement, mais assez lentement pour rester homogène et pour
    que la vitesse de rotation soit la même en tous ses
    points. D'abord presque sphérique, elle s'aplatit de plus en plus,
    en conservant la forme d'un ellipsoïde de révolution. Puis
    l'équateur lui-même cesse d'être circulaire et devient elliptique,
    la masse fluide prend alors la forme d'un ellipsoïde à trois axes
    inégaux. Ensuite l'ellipsoïde se creuse dans sa partie médiane
    l'une de ses moitiés tend à s'allonger de plus en plus et l'autre
    à se rapprocher de la forme sphérique. Enfin tout porte à croire
    que si le refroidissement continuait encore, notre masse se
    partagerait en deux corps distincts et inégaux.''  (Henri Poincaré,
    manuscript of an unsigned article published in \emph{Le Temps} on
    5 May, 1886, auctioned in Paris by ALDE, lot N°~292, on 6 May,
    2008, edited in \citet{hp2016cp},
    §~\href{http://henripoincarepapers.univ-nantes.fr/corresphp/index.php?a=on&id=4016}{3-48-1}.)
    All translations are my own, unless otherwise indicated.}
\end{quote}
The mathematical theory of the pear motivating Poincaré's thought
experiment had recently appeared in the pages of
\emph{Acta Mathematica}, along with an illustration (Fig.~1) comparing
a contour of his pear to that of a Jacobi ellipsoid, represented by a dashed line.

\begin{figure}
  \centering
  \includegraphics[width=80mm]{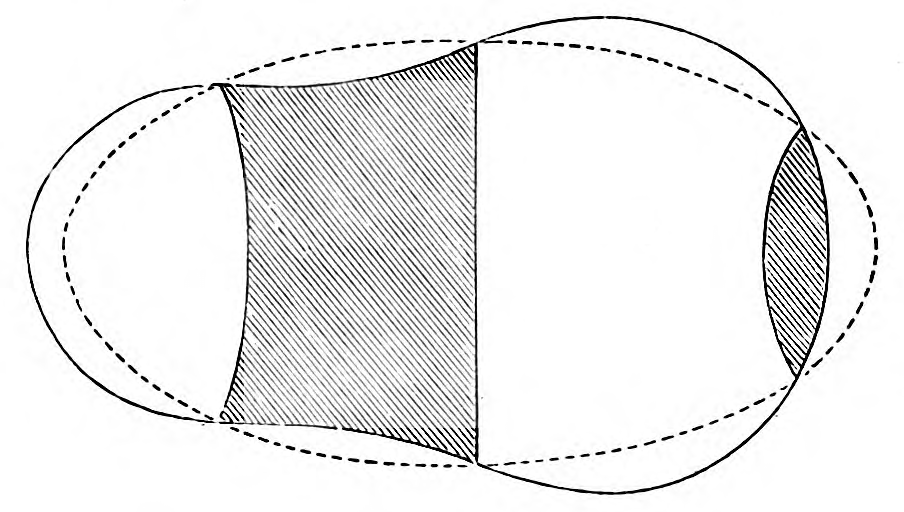}
  
  \caption{Poincaré's illustration of the piriform figure, from
    \citet{hp1885ama}.}
  \label{fig:pear}
\end{figure}

There are two aspects of Poincaré's discovery that I want to
underline. First of all, the identification of a new series of
equilibrium figures was a remarkable event in the history of
mathematics, placing Poincaré's name alongside that of Jacobi, and
giving rise to exchanges with Liapunov, William Thomson and George
Howard Darwin, among others.\footnote{For a gentle introduction to
  stability theory in the nineteenth century, with a focus on
  Poincaré, see \citet{ArcT2015}. Poincaré's contribution is discussed
  by \citet[chap.~5]{GraJ2013}.} Four decades later, Élie Cartan showed
that the stability of the pear could not be proven by Poincaré's
method \citep[4]{LytR1953}, which leads me to my second point.

From the outset, Poincaré stressed \emph{fission} as a possible outcome for a
homogeneous rotating fluid mass. A rotating fluid mass in space could then be
considered as a possible progenitor of a binary star, if one were to
admit stars are adequately modeled as homogeneous fluid masses. In the
state of knowledge of the internal structure of celestial objects in
the late nineteenth century, this was not an unreasonable supposition,
and it was one made readily by Poincaré and other
mathematical astronomers.

The fission theory of the origin of binary stars flatly contradicted
Laplace's nebular hypothesis, according to which the planets of the
solar system evolved (roughly) via condensation of diffuse matter
surrounding and rotating along with the sun. A similar scenario was
imagined for the formation of spiral nebulae, fueling speculation about the age of
the cosmos, and providing a mathematical foundation for the evolution
of the universe. With the emergence of the science of thermodynamics
in the 1850s, and the discovery of the second law of thermodynamics,
the universe was understood by physicists like W.~Thomson to
be approaching its demise, the so-called ``heat-death of the
universe''.\footnote{On the nebular hypothesis in astronomy and
  geophysics, see \citet{BruS1996b}; on its broader interpretation,
  see \citet{BeeG1989,SchS1989a}.}
Poincaré was fully aware of the contradiction, and he sought to
deflate it by noting that the nebular hypothesis concerned a
heterogeneous matter distribution, while the fission theory assumed a
homogeneous fluid mass \citep[379]{hp1885ama}.

The fission theory was associated first with Darwin, who was intrigued
by the evolution of orbital parameters of the
Earth-moon system, including the gradual decrease of the Earth's rotational
velocity. Taking into account tidal forces, and working backwards in
time, Darwin estimated that these two bodies resulted from fission no
less than fifty-four million years ago \citep[882]{DarG1880}.
As Darwin explained, he was originally motivated to study this problem
by the Kant-Laplace nebular hypothesis:
\begin{quote}
  It was in the hope that the investigation might throw some light on
  the nebular hypothesis of Laplace and Kant that I first undertook
  the work. It must be admitted, however, that we do not obtain much
  help from the results. It is justly remarked by M.~Poincaré that the
  conditions for the separation of a satellite from a nebula differ
  from those of his problem in the great concentration of density in
  the central body. \citep[442]{DarG1887a}
\end{quote}
Unlike Darwin, Poincaré worked forward in time, while assuming, like
Darwin, a homogeneous rotating fluid mass. Poincaré further developed
a powerful analytical approach to the study of equilibrium figures building on the
contributions of Jacobi, Joseph Liouville, and W.~Thomson and Peter Guthrie
Tait.\footnote{On Liouville's largely unpublished work on equilibrium
  figures of rotation, see \citet{LutJ1984}.}

A crucial concept in Poincaré's approach is that of a point of
bifurcation in a sequence of equilibrium figures arranged linearly
with respect to some parameter (e.g., angular velocity), or to a system
of parameters. He wanted to show that new equilibrium forms could, for
a given parameter, bifurcate from such a sequence, and exchange their
stability at the branch point \citep{MawJ2014}. James Jeans later
considered the two-dimensional case of a rotating infinite cylinder
\citep{JeaJ1903}, which features several analogies to the
three-dimensional case, and as Darwin underlined in his review of
research on the genesis of binaries, leads to a pear-shaped
equilibrium form \citep{DarG1909a}. The separate question of the perturbative
stability of the pear was of interest to mathematicians and
theoretical astrophysicists alike, in a context of competing definitions of
stability \citep{RoqT2011}.

Double stars had long captured the attention of astronomers, along
with variable stars. In the late nineteenth century, aided in
particular by more powerful telescopes, and new methods in
photography, photometry and spectroscopy, astronomers and
astrophysicists learned much about these objects, and began cataloging
them according to sky position, magnitude and spectral class. The
number of known binary stars grew rapidly when spectral lines of
certain stellar objects matched those of the spectra of two stars in
distinct spectral classes (or subclasses). Estimates of binary star
density in the general stellar distribution grew accordingly. In 1911,
Poincaré thought that one in three stars ``at least'' was binary
\citep[394]{hp1911rm}; William Wallace Campbell put this figure at one
in five or six stars, for the spectral classes F, G, K or M
\citep[280]{CamW1913}. Knowing something about the origin of binary
stars had become a primary objective for astronomy by the second
decade of the twentieth century.

Scientific interest in variable stars dates from the early eighteenth
century, when the observed variation in magnitude was attributed to
(unseen) dark spots, which would recur periodically with rotation of
the star. Motion of sunspots offered a concrete local example of this
phenomenon. During the final two decades of the century, the York duo
of Edward Pigott and John Goodricke -- the latter a young deaf-mute
gentleman -- discovered a number of new variables in Pigott's private
observatory. One of these was Algol ($\beta$ Persei), the sudden
change in magnitude of which prompted Pigott to conjecture, in 1781,
the existence of a dark eclipsing companion. News of this discovery
was hailed at the Royal Society of London, while William Herschel, who
had just discovered the first new planet in recorded history,
predicted that the hypothesis of ``a plurality of solar and planetary
systems'' would soon be verified \citep{HosM1979}. The Royal Society
lost no time in awarding Goodricke the Copley Medal for the discovery
of Algol's periodic variation in luminosity. In 1786 Goodricke, at the
age of 21, and just two weeks before he succumbed to pneumonia, was
named a Fellow of the Society.

After the triumph of their discovery of Algol's periodicity, Pigott
and Goodricke naturally continued to seek out new variables,
and over the next few years they found three more: $\eta$ Antinoi (now
$\eta$ Aquilae), $\beta$ Lyrae and $\delta$ Cephei. They recognized
that the luminosity variation of these stars did \emph{not} lend itself so
immediately to an interpretation in terms of an eclipsing
companion. The first and last of the trio are in fact Type I Cepheid
variables. As for $\beta$ Lyrae, Goodricke correctly characterized its
variation over 12.91 days, with \emph{two} maxima of equal luminosity
and \emph{two minima}.

The theoretical upshot of the latter discoveries was to render quite
doubtful the eclipse hypothesis, even for the explanation of Algol's
variation. Pigott seems to have soured on the eclipse theory, as he
went on to adopt the view of W.~Herschel, according to which
observed variation in magnitude results from the rotation of a single
dark body covered by a luminous atmosphere of varied thickness. This
theory did not invite geometrical modeling at first, and the eclipse
hypothesis went dormant for most of the nineteenth century.

Variable stars held intrinsic interest for a few observers in the
early nineteenth century, foremost among whom was the Bonn astronomer
Friedrich Wilhelm August Argelander. He observed and published the periods of hundreds
of variables, including that of $\beta$ Lyrae, and inspired others to
do likewise. Many of Argelander's observations were cataloged in
Johann Carl Friedrich Zöllner's \emph{Photometrie des Himmels} \citep{ZolF1861},
which carefully distinguished known variables.\footnote{The Leipzig
astrophysicist Zöllner is often credited with the foundation of
astrophotometry, in light of his design in 1861 of a photometer with a
flame reference and a pair of Nicol prisms as the polarizing apparatus
\citep[385]{PanA1961}.}

Zöllner's photometric observations, and his revival of W.~Herschel's
rotation theory of variables were a source of inspiration for the
Swedish mathematical astronomer Hugo Gyldén. In 1880, Gyldén took up a
special case of the rotation theory of variables, by placing
constraints on the light curve produced by a body of non-uniform,
constant surface luminosity in uniform rotation about its principal
axis of inertia. Gyldén offered no numerical results, and mentioned
only two specific variables: Mira Ceti and $\beta$ Lyrae. The latter
of these, Gyldén wrote, was of ``little interest from a theoretical
standpoint,'' because its axis of rotation nearly coincided with its
inertial axis, ``as far as this can now be seen''
\citep{GylH1880c}. In general, until the late 1880s, variables failed
to inspire mathematical astronomers.

Visual binary stars, on the other hand, were of some interest to
mathematical astronomers. Mathematicians worked with two
ellipses: the true path of a companion orbiting the principal
star, and its projection on a plane orthogonal to the line of sight,
referred to as the apparent ellipse.\footnote{In the
1890s, astronomers referred to these as the ``relative'' and
``absolute'' orbits, respectively.} A transformation
was then sought from the former to the latter. Complications set in
with the sensitivity of observations of motion at apparent aphelion
and perihelion, which John Herschel (the son of William) sought to manage with
probabilistic considerations.\footnote{\citet{HerJ1850a}. On the
  nature of astronomers' interest in double stars, see
  \citet{WilM1984}, and on J.~Herschel in particular, see \citet{CasS2018}.}

The number of known binary stars increased rapidly with the help
of spectroscopy. At the Potsdam Observatory, Hermann Carl Vogel (1841--1907)
and Julius Scheiner discovered their first \emph{spectroscopic} binary in 1889
\citep{VogH1890}. The binary star they observed was one of the
variable stars discovered by Goodricke; Vogel noted that the
displacement of spectral lines of the brighter component matched the
phase of Algol's light curve. Two other spectroscopic binaries were
discovered at roughly the same time at Harvard College Observatory
(HCO) by Edward C.~Pickering and Antonia
Maury \citep[53]{HeaJ2014}. In 1893, Vogel was awarded the Henry
Draper Medal by the National Academy of Sciences (USA), ``for
spectroscopic observations upon the motion of stars in the line of
sight, and other kindred researches.''

Vogel was a pioneer of spectroscopic measurement of radial velocity,
and a champion of research on variable stars. His analysis in 1894 of
recent observations of $\beta$ Lyrae declared this variable to be
\begin{quote}
[..] among the most interesting spectroscopic objects in the northern
heavens.\footnote{The original sentence reads: ``Der
  veränderliche Stern $\beta$ Lyrae, bemerkenswerth durch die
  eigenthümliche Form seiner Lichtcurve, zählt in spectralanalytischer
  Beziehung zu den interessantesten Objecten des nördlichen Himmels''
  \citep{VogH1894}. Vogel's paper was translated to English; see
  \citet{VogH1894a,VogH1894b}. The given translation is borrowed from
  the latter publication.}
\end{quote}
Vogel's paper took up the spectrograms realized at the Pulkovo
Observatory by Aristarkh Belopolsky, and correlated graphically to Argelander's
light curve. Belopolsky interpreted the variable as an partially-eclipsing
binary with a circular orbit \citep{BelA1894}. While Vogel did not
doubt that $\beta$ Lyrae is a close binary star, he noted that the
light curve was equally consistent with an elliptical orbit, where the
major axis is aligned with the line of sight. The real difficulty for
Belopolsky's interpretation was the following: at the minima of luminosity, the
relative radial velocity of both components must vanish. Similarly, for the
luminosity maxima, which ought to correspond to the moments of maximum
relative line shift, in opposite directions for the two maxima. None of this,
however, was confirmed by Belopolsky's spectrograms. Vogel
emphasized the difficulty of reconciling photometric and spectroscopic
observations of $\beta$ Lyrae. Astronomers love a good challenge, and
they were well served here by Vogel. Determining the orbital parameters
for a system compatible with photometric and spectroscopic
observations, and with other constraints from physical considerations
became what Stein later called the ``$\beta$ Lyrae problem''
\citep[379]{SteJ1924}.

Along with Vogel, Pickering was another decisive figure in the rise of
variable-star research. As HCO director, Pickering published the first
spectroscopic study of Algol, realized from photographs taken
over four years, in collaboration with Williamina Fleming and Antonia Maury. The
spectrum of $\beta$ Lyrae, Pickering announced, ``is unlike that of
any other star hitherto examined'' \citep{PicE1891b}. He interpreted
the recorded line shifts as the signs of a close binary with
components of unlike spectra, but he was open to two other
possibilities. One of these was a ``meteor stream'', understood to be
an eclipsing swarm; more on this will follow below. The other
possibility was a single rotating sun-like star with
a large protuberance which, due to the observed periodic doubling of
lines, had to extend beyond the stellar equator.

The discoveries by Vogel, Pickering and others put the eclipse hypothesis back
into play for the explanation of short-period variables. A variant of
sorts of Pigott's eclipse hypothesis was put forth at this time by
Norman Lockyer (1836--1920), the self-taught astronomer, founder and
editor of \emph{Nature}, and from 1877, first director of the Solar Physics
Observatory in South Kensington \citep[53]{HeaJ2014}. Lockyer supposed short-period variables to be
eclipsing, interpenetrating meteor swarms, where the heat of collision
gives rise to vapor. In fact, this was just the latest avatar of the
all-embracing ``meteoritic hypothesis'' he had advanced in 1887,
and from which he derived his scheme of stellar evolution. The scheme
evolved over time, but the basic idea remained the same: stellar
evolution followed a temperature curve arranged according to rising or
falling temperature, with the young stars increasing in temperature,
changing spectral class along the way, reaching a maximum, and
decreasing in temperature with advancing age. Lockyer's
friend George Gabriel Stokes found the meteoritic hypothesis to beg the question of the
origin of crystals in meteors, but as an explanation of short-period
variables, the image of eclipsing meteor swarms had a degree of
plausibility. Pickering found it useful for understanding the line
shifts of $\beta$ Lyrae, as noted above.\footnote{Under Lockyer's hypothesis, collision of
  meteors occurs between two or more companion swarms, and also
  between the latter and swarms passing through
  interstellar space; for a more complete presentation, see \citet[200]{MeaA2008}.}

\begin{figure}
  \centering
  \includegraphics[width=80mm]{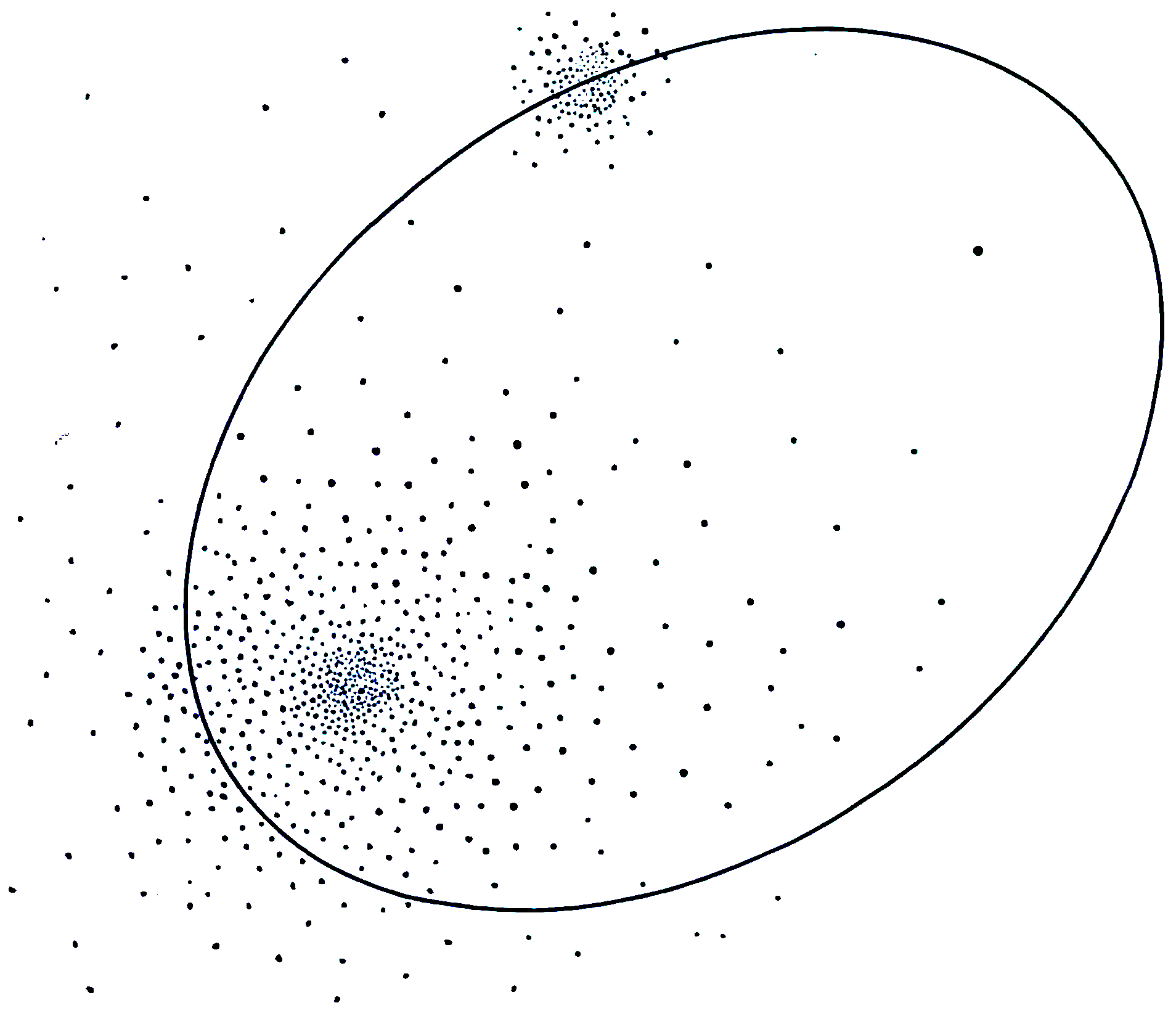}
  \caption[Lockyer's swarms]{Illustration of two luminous meteor swarms in
    Keplerian orbit \citep[479]{LocJ1890}.}
  \label{fig:swarms}
\end{figure}

\begin{figure}
  \centering
  \begin{tabular}{ccc}
    \includegraphics[width=30mm]{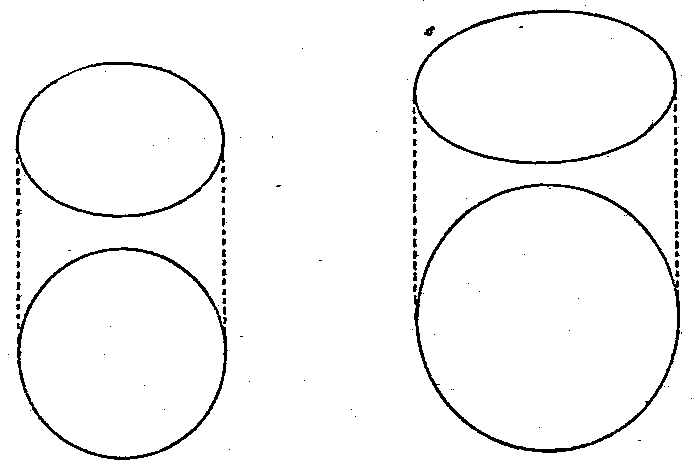}&
  \includegraphics[width=35mm]{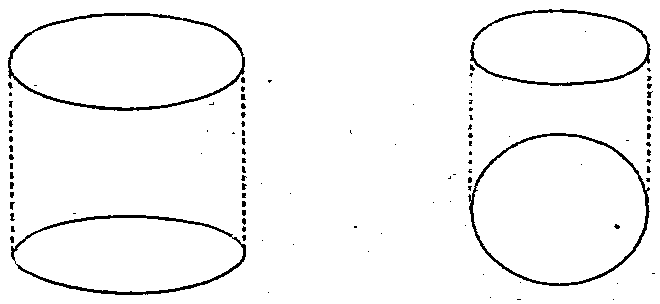}\qquad&\qquad                                             
  \includegraphics[width=35mm]{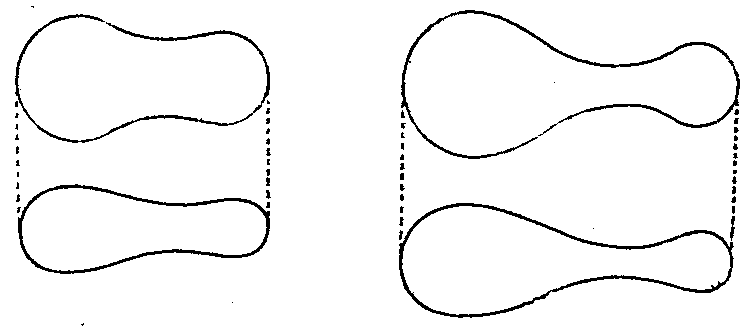}\\
    a.~Ellipsoid of revolution   & b.~Jacobi ellipsoid  & c.~Pear
  \end{tabular}
  
  \medskip
  \includegraphics[width=35mm]{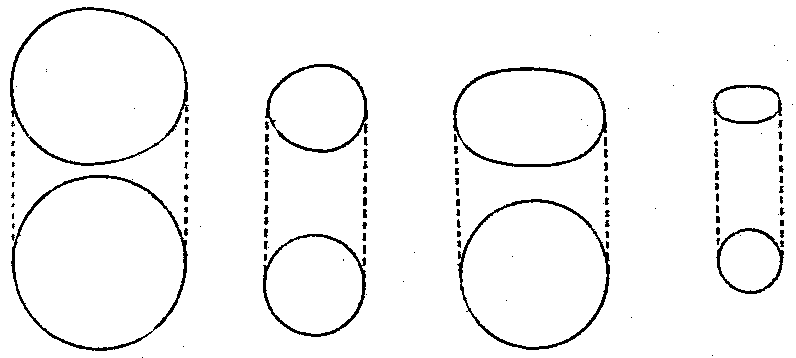}

  d.~Post-fission pair of pears
  evolving into a pair of ellipsoids
  \caption[Equilibrium figures]{Evolution of an equilibrium figure,
    from \citet{hp1892rga}. The temporal evolution of a single
    rotating fluid mass is illustrated from (a) to (d). Stacked
    figures are orthogonal projections, such that the axis of rotation
    points out of the page in the lower figure. In each case, the left
    figure evolves into the right figure.}
  \label{fig:hp1892rga}
\end{figure}

Poincaré, too, contributed to the rise of the eclipse hypothesis, both
directly and indirectly. In 1892, he published a general account of
equilibrium figures, featuring a series of contours illustrating the
evolution of a rotating fluid mass which culminates in fission (Figure
\ref{fig:hp1892rga}). As for the \emph{physical reality} of such an
evolution, Poincaré drew a direct link to close binaries:
\begin{quote}
  Perhaps the process I have just described [..] is closer to the one that
  produced certain double stars than that from which the solar system
  emerged. All this remains very hypothetical in any
  case.\footnote{``Peut-être le processus que je viens de décrire
    [..] se rapproche-t-il plus de celui qui a produit
    certaines étoiles doubles que de celui d'où est sorti le système
    solaire. Tout dans tous les cas reste très hypothétique.'' \citet[813]{hp1892rga}}
\end{quote}
Poincaré directed the attention of his readers to the
deployment of his fission hypothesis beyond the solar system, while
not ruling out its use closer to home. As we shall see in what
follows, over the next ten years,
Poincaré revised his view, and came to regard fission of a
self-gravitating, rotating star
as the most probable explanation of binary formation.

We can chart the growing interest in variables from 1888 to 1896 based
on Seth Chandler's three catalogs. In his presentation of the
catalogs, Chandler insisted on the rigorous nature of his selection,
while promoting additional discovery efforts. For example, the third
catalog reports the recent discoveries of a total of 153 variables in
seven globular clusters by Solon Bailey at the Arequipa Observatory,
confirmed by Pickering and Fleming. Chandler omitted these newly-discovered variables
from his third catalog, as he did not yet have all the details he
required, and he did not wish to delay
publication \citep{ChaS1888,ChaS1893,ChaS1896}.
  The task of cataloging new
variables soon became overwhelming, and Chandler decided to hand it
over to the \emph{Astronomische Gesellschaft}. This was surely a wise
move on his part, as the announced discovery of variables in clusters was only the
beginning of a torrent of discoveries. Annie Jump Cannon's catalog of
1907 featured 1425 variables, of which roughly a third were found in
globular clusters \citep[IV]{MulG1918}.

\begin{figure}
  \centering
  \includegraphics[width=80mm]{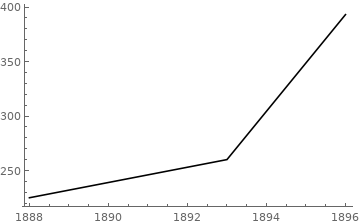}
  \caption[Variable-star counts]{Discovery rate of variable stars, from
    S.~C.~Chandler's three catalogs, 1888--1896.}
  \label{fig:chandler}
\end{figure}

\section{G. W. Myers' models of $\beta$ Lyrae and U Pegasi}

The first mathematical astronomer to pick up Vogel's challenge was a
young American: George William Myers
(1864--1931). Largely forgotten in the general history of astronomy,
if not to \citet{KuiG1941}, \citet{StrO1958} and others who study variables,
Myers studied engineering at the University of Illinois.  Like many
other American students in the exact sciences at the time, he went to
Germany to obtain a Ph.D.\footnote{For an overview of the emergence of
  mathematical research in the United States, see \citet{ParK1994}.} In 1896, he defended a thesis in
theoretical astronomy at the University of Munich on the variability of $\beta$
Lyrae.\footnote{From 1888 to 1900, Myers held several positions in
  succession at the University of Illinois, from instructor of
  mathematics to assistant, associate, and full professor of
  mathematics and astronomy \citep[885]{Marq1943}. Upon completion of his thesis in Munich,
  he directed the new observatory in Urbana. In 1900, Myers became
  head of astronomy and mathematics at the Chicago Institute, an
  institution for educating future schoolteachers founded by Francis
  Wayland Parker (1837--1902), a pioneer of the progressive school
  movement in the United States. In 1901, the Chicago Institute became
  the School of Education of the University of Chicago 
  \citep{DewJ1902}, where Myers was named professor of mathematics
  education and astronomy, a position he held until his retirement in
  1929. Myers was a member of the astronomical societies of Germany,
  France, Belgium, Mexico and South America, and belonged to the
  American Academy for the Advancement of Science, and the American
  Mathematical Society.}

Myers' thesis advisor, and director of the Munich Observatory, Hugo
von Seeliger suggested that he work out $\beta$ Lyrae's orbit. The
choice of topic was likely inspired in part by Seeliger's own
career. As a young man, Seeliger wrote his thesis in Leipzig on the
orbits of binaries, and was employed from 1873 to 1878 by Argelander
at the Bonn Observatory \citep[4]{WilA1927}. More recently, Vogel's
study had shown the $\beta$ Lyrae problem to be both interesting and
challenging, perhaps even impossible. Seeliger had previously charged
another doctoral student, Carl Harting, with the construction of a
model of the orbit of Algol \citep{HarC1889}. And while
\citet{PicE1880c} made progress with the latter star based on its
light curve, and found it to be an eclipsing binary, Harting was
the first to determine a differential light curve from a linear
variation of orbital parameters \citep[283]{SteJ1924}. Of course,
Harting did not take into account any measurements of radial velocity,
as these were only obtained for Algol the year after his
thesis defense.

The spectrographic method employed by Pickering, Vogel, Maury and
others held real promise as a tool for discovering binary stars, and
for providing the elements needed to calculate their orbit. In the new
class of spectroscopic binaries, a subclass of variables emerged, and
soon engulfed all variable stars, as variables were generally assumed
to be eclipsing binaries. This remained true even as late as 1918, as
Robert Aitkens' monograph shows. At the University of Illinois
Observatory, Joel Stebbins, for example, expressed the common-sense
view to the effect that ``all spectroscopic binary stars would be
eclipsing variables for observers properly situated in space''
\citep{SteJ1911}.

By the mid-1890s, several astronomers had produced light curves for
the variable $\beta$ Lyrae. Myers used one with 1439 measurements
realized over a span of nineteen years by Argelander, and published in
1859.  The idea for Myers was the same as for Vogel: to tune the
parameters of his geometric model to obtain a close representation of
both the spectroscopic velocity-curve and the photometric
light curve. As established by Argelander, $\beta$ Lyrae's luminosity
varied continuously over a period of about 12.91 days, with two
roughly-equal maxima and two unequal minima. Myers reasoned that
orbiting spheres would never deliver the target light curve, and so he
worked with two ellipsoids, one a satellite in circular orbit, and
both revolving with their major axes in perfect alignment. When the
major axes of the true system are perpendicular to the line of sight,
we then have the two maxima. From such geometric relations, Myers
determined the ratio of respective ellipsoid axes that best fit the
light curve at maxima. He then iterated the orbital elements
(eccentricity, longitude of periastron, inclination with respect to
line of sight) until he obtained a close fit. The result of Myers'
labors was a very close binary, where the distance of the ellipsoid
centers is only about 2.4 times the semi-major axis of the
larger ellipsoid \citep{MyeG1896,MyeG1898}.

\begin{figure}
  \centering
  \includegraphics[width=55mm]{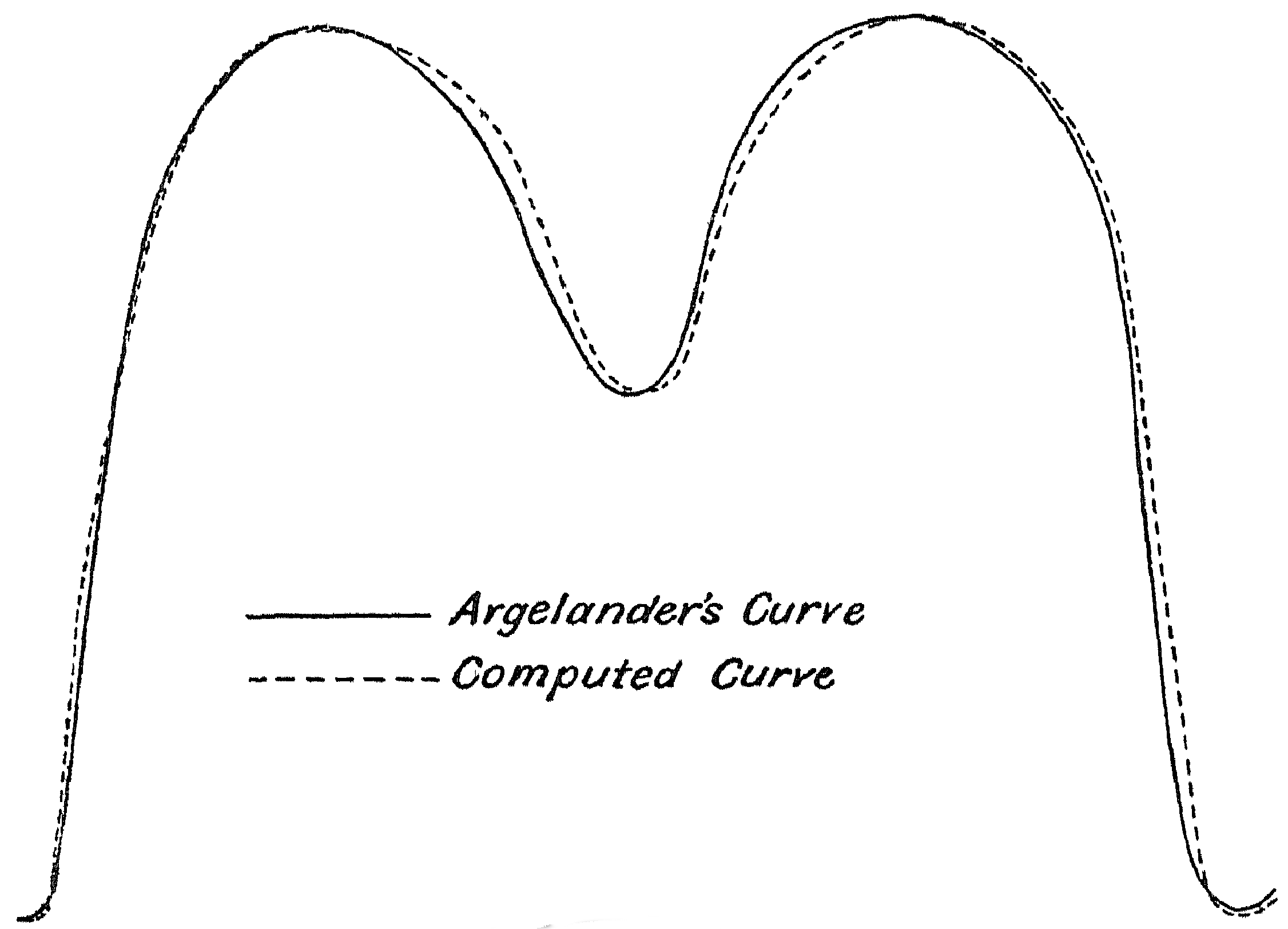}\qquad
  \includegraphics[width=55mm]{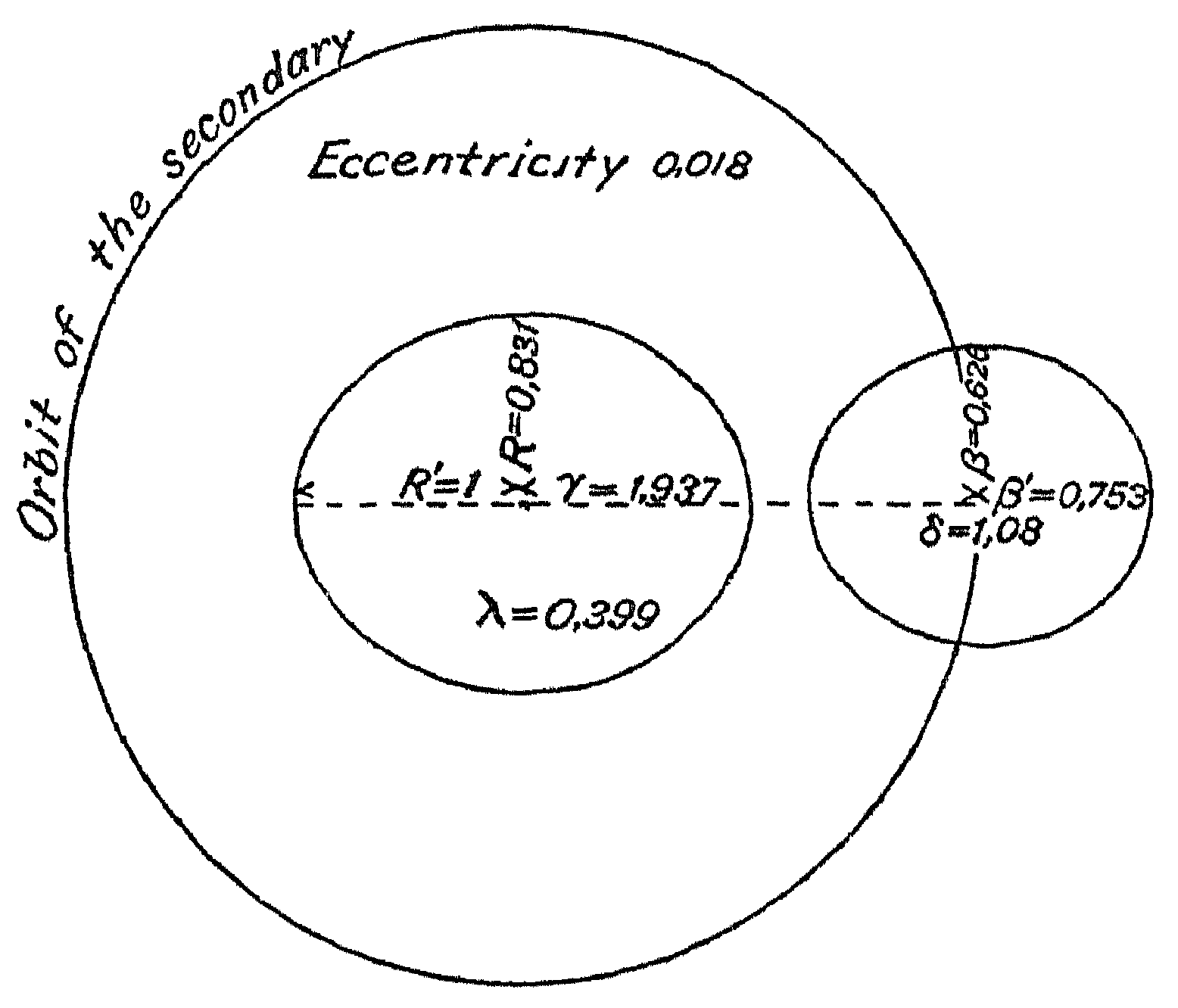}
  \caption[Myers' orbit]{Observed and theoretical light curves, and
 theoretical orbit of $\beta$ Lyrae, from \citet{MyeG1898}. Orbital
 eccentricity is small, and the axes of the ellipsoids are always aligned.}
  \label{fig:myers}
\end{figure}

In order to convince himself of the correctness of his method, Myers
explained, he perturbed an orbital element, and computed the
corresponding change to the light curve. By doing so, he noticed that
an orbital solution for which the ellipsoids \emph{interpenetrated} produced a
light curve with ``nearly the same'' probable error as that of his
final result. From this circumstance, Myers concluded that $\beta$
Lyrae was either ``not yet separated'', or of recent separation, and
consequently,
\begin{quote}
  \dots we seem to have here the first concrete example of a world in
  the act of being born. \citep[17]{MyeG1898}
\end{quote}

Having concluded the task of orbital determination from the
light curve, Myers went on to consider the spectrograms obtained by
Belopolsky for $\beta$ Lyrae in 1892 at the Pulkovo Observatory
(mentioned above). Myers established a radial velocity curve from
these spectrograms, from which the orbital elements were readily
determined using the method Arthur \citet{RamA1891} successfully
applied to Pickering's observations of the binary $\beta$
Aurigae.\footnote{Useful overviews of models of the motion of binary
  stars are provided by \citet{HepJ1910,AitR1918,HenF1928a}.}

In order to determine the inclination of the orbit, Myers reasoned
that from the form of the light curve, the ellipse had to lie on a
plane intersecting the sun. The absolute orbit was then fixed, and
Myers proceeded to work out the relative orbit. To do so, Myers relied
on an observation of the relative displacement of three dark lines communicated
privately by Lockyer, which he converted to velocities, and corrected
for terrestrial motion. By calculating the corresponding relative
velocities from the absolute orbit, Myers found the ratio of absolute
to true velocities
to be 1/3.168, which is just the ratio of major axes, $a/A$, and also
that of the two masses, $m/(m + M)$. Since Myers had no way of
knowing which of the masses, $m$ or $M$, produced Belopolsky's lines,
he made an educated guess, such that the mass of the smaller star was
about 9.5 solar masses and the larger, about 21 solar masses.\footnote{Myers' value
  for the system mass of $\beta$ Lyrae is
  high by current estimate, which puts it at roughly
  sixteen solar masses (SIMBAD).} From the light curve analysis,
Myers knew the volume of the ellipsoids, and he estimated the mean density
of the system to be less than that of air. This indicated a
nebular system, which prompted Myers to comment:
\begin{quote}
  It appears then that $\beta$ Lyrae furnishes us a concrete
  illustration of the actual existence in space of a Poincaré figure
  of equilibrium. \citep[19]{MyeG1898}
\end{quote}
Remarkably, Myers concluded his thesis with the observation that
$\beta$ Lyrae was not a \emph{binary} system at all. Instead, the
combined photometric and spectroscopic observations of the system
indicated that it was a \emph{single} spinning body -- a body whose
form, he wrote, had been studied mathematically by Poincaré and
Darwin.

After the successful defense of his thesis in Munich, Myers returned
to Urbana, where he was promoted to full professor of astronomy and
mathematics, and director of the new observatory.\footnote{Myers directed the
University of Illinois Observatory until 1900, when he moved
to the Chicago Institute \citep[885]{Marq1943}.} On October 20, 1897, he
presented his theory of $\beta$ Lyrae to astronomers gathered in
Williams Bay, Wisconsin, for the
inauguration of the Yerkes Observatory. His paper, entitled ``The
system of $\beta$ Lyrae'', appeared in the recently-founded \emph{Astrophysical
  Journal}, which assured its diffusion among astrophysicists. In his paper's conclusion, Myers reiterated and
strengthened his interpretation of $\beta$ Lyrae as a spinning pear:
\begin{quote}
  In conclusion, let it be observed that an attempt at a formal
  representation of the condition of things prevailing in the system
  of $\beta$ Lyrae, leads to the assumption of a single body (such as
  Poincaré's or Darwin's figures of equilibrium). The above has, of
  course, only a formal significance, but on account of the poverty of
  observational material at my disposal an attempt to push the
  discussion farther on a mathematical basis could not have proved
  profitable. It is believed, however, that the discussion may help us
  to orient our views with regard to this wonderfully interesting
  star. \citep[22]{MyeG1898}
\end{quote}
Myers took care here to associate his model with the pear, and to
suggest that further work was required, both from mathematicians and
observers.

The inauguration of the Yerkes Observatory, organized by its director,
George Ellery
Hale, was arguably the astronomical event of the year for North
American astronomy. In attendance for the three-day meeting that
preceded the inauguration were sixty-odd physicists and astronomers, in
a Who's Who of turn-of-the-twentieth-century astronomy and astrophysics, including
the Göttingen spectroscopist Carl Runge, James E.~Keeler (who
delivered the keynote address), Albert A.~Michelson, George W.~Ritchey
and Forest R.~Moulton. Myers' talk was featured in an afternoon
session, along with others by Simon Newcomb, Pickering, Edward E.~Barnard,
George W.~Hough and Father Hagen.\footnote{Yerkes Observatory program,
  University of Chicago Library, Special Collections Research Center;
  undated, page 4. On the inauguration, see \citet{OstD1999}. At least
  eighty people were in attendance for the inauguration; many were
  captured in a photo reproduced in \citet{OstD1997}.}
The publicity afforded Myers' results on $\beta$ Lyrae by the
Yerkes meeting, and their subsequent publication helped consolidate
acceptance of the Poincaré pear and Poincaré-Darwin fission theory in astrophysics.

In the wake of the Yerkes inauguration, Pickering, impressed
by Myers' results, asked O.~C.~Wendell to establish a new light curve for U
Pegasi. Originally thought to be an Algol-type eclipsing binary with a
very short period of 4.5 hours, this classification was contested, and
Pickering hoped to settle the matter with new photometric data.\footnote{Pickering
  noted Chandler's suggestion that U Pegasi represented a new class of
  variable, distinct from both the Algol type and that of the short
  period variables $\eta$ Aquilae and $\delta$ Cephei
  \citep{PicE1898}.} On 28 December, Wendell recorded two unequal minima,
such that the period of the star was doubled, and the resulting light curve of
U Pegasi, Pickering wrote, ``closely resembled that of $\beta$ Lyrae''
\citep{PicE1898}.

Pickering did not attempt to work out the orbit of U Pegasi, but
instead invited Myers to do so, extending him the use of the HCO
photometer. Once he had verified the difference in the observed
magnitude in minima found by Wendell, Myers applied his method to
determine the probable orbit of U Pegasi from the HCO light
curve. This was somewhat simpler than calculating the orbital
parameters of $\beta$ Lyrae from Argelander's light curve, in that
satisfactory results were obtained from an assumption of eclipsing
spheres, without having to resort to ellipsoids of revolution.

The result of Myers' analysis of U Pegasi's light curve will come as
no surprise, given Pickering's identification of it (in January, 1898)
as a $\beta$ Lyrae class variable. Pickering invited Myers to present his
results at the Harvard Observatory
Astronomical Conference held in the drawing room of the director's
residence on 19 August, 1898.\footnote{Among the 93 registered
  attendees of the ``Second
  Annual Conference of astronomers and astrophysicists'' were Hale, Newcomb,
Stein (visiting from Leiden), Charles St.~John, Benjamin Peirce and
the HCO staff; most attendees were from the northeast
coast of the United States. For the full attendance list of the three-day meeting, see \citet{HalG1898}.} Myers announced that,
not only is U Pegasi's light curve ``satisfactorily represented by the
satellite theory'', but in addition, it is probably a pear:
\begin{quote}
  The distance of centers does not materially differ from the sum of
  the radii of the components, suggesting the probable existence of
  the ``apiodal'' form of Poincaré. \citep[172]{MyeG1898a}
\end{quote}
With much encouragement from the HCO director, Myers had now tentatively
identified a second variable as a Poincaré pear.\footnote{Later studies would
bear out the general classification, although U Pegasi was eventually
reclassed as an eclipsing binary of a third type, unknown
in 1898: W UMa. For a more recent study, see \citet{DjuG2001}.}

Myers was not the only one to endorse the pear, and to latch on to the
fission theory: Karl Schwarzschild, Myers' fellow doctoral student in
Munich did the same. Schwarzschild defended his doctoral thesis under
Seeliger's direction in 1896, on the topic of Poincaré's theory of
rotating fluid masses \citep{SchK1898}. Much like Myers, Schwarzschild
expressed his readiness to tread where Poincaré would not, but his
reasoning followed a different path, one indicated by the branching sequence of equilibrium forms:
\begin{quote}
  We don't yet know exactly what is happening here. Probably, as
  chance will have it, a smaller part of the mass emerges from the
  ellipsoidal form at one end or the other of the major axis, and the
  whole mass assumes the pear-shaped form of Poincaré's figures, after
  which there will be an ever-stronger indentation, and presumably in
  the end a splitting of the mass into two unequal parts. [..]
  Mr.~Poincaré considers it too bold to want to infer from this
  history of an invariably-homogeneous mass the reform of Laplace's
  inhomogeneous nebula. However, if one were to imagine not a gaseous
  mass, but a liquid one which, even with vanishing surface pressure,
  always has finite density, [..] then one may conclude that such a
  liquid mass undergoing increasing contraction will lose its
  rotational form, and eventually split.\footnote{``Was hier geschieht, wissen wir noch nicht genau. Wahrscheinlich tritt ein kleinerer Teil der Masse, wie es wieder der Zufall will, am einen oder andern Ende der grossen Axe aus der ellipsoidischen Form heraus, und die ganze Masse nimmt die birnförmige Gestalt der Poincaré'schen Figuren an, worauf eine stärkere und stärkere Einkerbung und vermutlich zuletzt eine Spaltung der Masse in zwei ungleiche Teile erfolgen wird. [..]
    Herr Poincare
    hält es für zu gewagt, aus dieser Geschichte einer stets homogen
    bleibenden Masse auf die Umgestaltung des von vornherein
    inhomogenen Laplace'schen Nebels schliessen zu wollen. Denkt man
    aber nicht an eine Gasmasse, sondern an eine Flüssigkeit, die auch
    bei verschwindendem Druck an der Oberfläche stets eine endliche
    Dichte behält, wie sie die Erde zu einer gewissen Epoche gewesen
    sein mag, so darf man folgern, dass auch eine solche Flüssigkeit
    bei zunehmender Kontraktion einmal die Rotationsform verlieren und
    sich schliesslich spalten wird.''  \citep[296]{SchK1898}}
\end{quote}
Like Poincaré, Schwarzschild viewed the pear as a binary precursor,
and found Poincaré's fission theory to be preferable to Laplace's
nebular hypothesis as an explanation of binary genesis, at least in
the absence of certain knowledge of the internal structure of
stars. The latter topic would interest Schwarzschild and others in the
coming decade, but at the turn of the century, this was not
yet an active field of research in astrophysics.

\section{Charles André's \emph{Traité d'astronomie}}

The most extensive review of Myers' theory of $\beta$ Lyrae appeared in French, in the
second volume of Charles André's \emph{Traité d'astronomie}. A former
student of the \emph{École normale supérieure} who was certified by
the \emph{agrégation} in physics, André (1842--1912) was employed by
the Paris Observatory in 1865 as an \emph{aide-astronome}. He was not
appreciated by the director, Le Verrier, but was recruited to a chair
in physical astronomy by the University of Lyon in 1877, and named
director of the newly-constructed observatory in 1879
\citep{VerP2016}.

In Lyon, André created a section for the study of variable stars in 1898, and
charged Michel Luizet with its operation \citep[8]{LuiM1912}. He also
lectured on recent research on variables, and wrote a two-volume
textbook on the subject, published in 1899 and 1900. In the preface to
the first volume, André wrote that his ``main objective'' in
publishing his lectures, was to
\begin{quote}
  \dots contribute to a return to favor, in our country, of the branch
  of observational astronomy [\dots] that, for various reasons, is a bit
  neglected at present \dots.\footnote{``\dots mon but principal est
    de contribuer à remettre en faveur, dans notre Pays, cette branche
    de l'Astronomie d'observation [\dots] qui, par des causes diverses,
    y est actuellement un peu délaissée \dots''. \citet[V]{AndC1899}}
\end{quote}
The idea that astrophysics was neglected in France was shared by
many, including Henri Poincaré.\footnote{On the social history of
  astrophysics in France, see the theses by \citet{LeGS2007} and
  \citet{SaiA2008}.}

Before discussing Poincaré's view of French astrophysics, a few words
about his bibliography are in order. Poincaré published over 700
titles in his lifetime, including several
prefaces, and reviews of research papers, but only two book
reviews. His review of the first volume of André's \emph{Traité} was
his first, and his review of André's second volume was the last he would ever
write. One wonders what drew him to this particular work?

Perhaps Poincaré took on the review, published in the journal he
directed, because like André, he felt that French astrophysics was not
keeping pace with international developments, and that André's
treatise could help fortify this emergent discipline in France. Poincaré
agreed with André's estimation that astrophysics was neglected in
France, and applauded him not only for the ``very great service to students''
that the volume represented, but also for ``attracting attention to
these questions so mysterious, so grandiose and so endearing''
\citep[124]{hp1899ba}.\footnote{Maurice Hamy, an astronomer at the
  Paris Observatory, also reviewed André's lectures, which he
  described as ``first-rate''. He agreed with
  André's assessment that his topic was ``ignored almost completely''
  in France \citep{HamM1901}.}

There was certainly another reason for Poincaré's review, as it was
the occasion for him to express his views on current trends in
astrophysics, and in particular, on statistical astronomy. The
possibility envisaged by André of determining the distribution of the few stars with
known parallax remained for
Poincaré an excessively speculative venture, because ``one is always reduced to
adventurous statistical data''.\footnote{``\dots on est
  toujours réduit aux données aventureuses de la Statistique.'' \citet[126]{hp1899ba}}

The book review was also the occasion for Poincaré to correct an error. A variable star,
André reasoned, whose light curve shows a periodicity that is itself
variable, allows us to discern its radial velocity, provided that the
variation is on the order of a ten-thousandth part of its
period. Poincaré was much concerned at this point of his career with the principle of
relative motion, with respect to the electron theories of Joseph
Larmor and H.-A.~Lorentz, and
Hertz's electrodynamics of moving media.\footnote{Poincaré's view of
  the principle of relative motion and the principle of reaction in
  1900 is explained in detail by \citet{DarO2023}.} He noticed that,
contrary to André's reasoning, under uniform motion, the dynamics of
an eclipsing binary can not change. In order for the period to change,
Poincaré pointed out, the variable star would have to undergo radial
acceleration. This is true, as far as it goes, but neither Poincaré
nor André entertained here the possibility that variables are not in a
state of equilibrium, although André went on to make such a suggestion
in the second volume of
his treatise.\footnote{Both $\beta$ Lyrae and U Pegasi
  were later considered to exhibit mass transfer.}

\begin{figure}
  \centering
  \includegraphics[width=55mm]{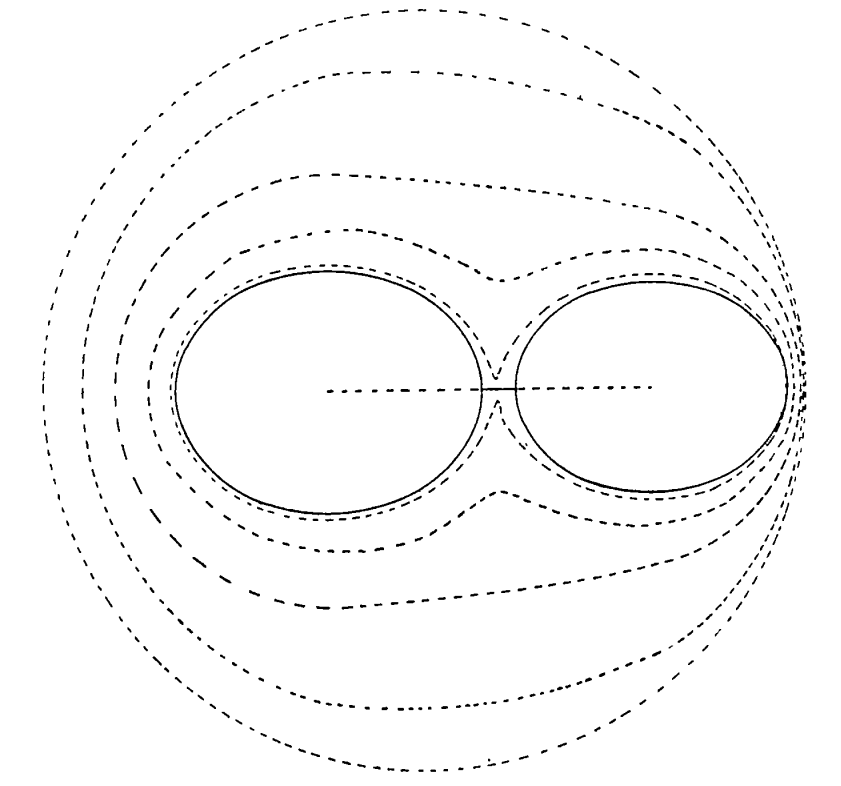}
  \caption[Equipotential surfaces]{Equipotential surfaces of Myers'
    model of $\beta$ Lyrae with two orbiting ellipsoids, featuring an
    apiodal surface \citep{MyeG1898}.}
  \label{fig:myers-equipot}
\end{figure}

The second volume of André's \emph{Traité d'astronomie} is devoted to
double stars, multiple stars and globular nebulae. André takes care to
acknowledge and repair the error in the first volume pointed out by Poincaré.
Among the double stars André takes up in this volume, $\beta$ Lyrae is
the most prominent, as André provides a detailed résumé of Myers'
model, and he agrees fully with the interpretation stressed by Myers in
Williams Bay:
\begin{quote}
  The binary system $\beta$ Lyrae constitutes \dots an absolutely
  remarkable case: \dots its two components are nearly in contact with
  each other. \dots The system U Pegasi is probably quite analogous,
  but with a closer contact of the two components. With such low mean
  densities, these two systems are probably still in a nebulous state;
  their atmospheres are indistinct and, intermixing at a certain
  distance from the stellar core, are distributed in equipotential
  surfaces differing more and more from a spherical form and taking
  on the forms so well studied by the eminent geometer Poincaré.\footnote{``Le système binaire de $\beta$ Lyre constitue \dots un cas
  absolument remarquable: \dots ses deux composantes sont presque
  en contact l'une avec l'autre.
  Il est d'ailleurs probable que le système de U Pégase lui est fort
  analogue, mais avec un contact plus assuré des deux composantes.
  Avec d'aussi faibles densités moyennes, ces deux systèmes sont
  probablement encore à l'état nébuleux; leurs atmosphères se
  confondent et, se mêlant à une certaine distance des noyaux
  stellaires, se distribuent en surfaces équipotentielles différant de
  plus en plus de la forme sphérique et affectant les formes si bien
  étudiées par l'éminent géomètre Poincaré.'' \citet[303]{AndC1900}}
\end{quote}
The equipotential surfaces were drawn by Myers in his Williams Bay paper;
André reproduced them in turn for his readers (Figure \ref{fig:myers-equipot}).

Poincaré appears to have been pleased with the second volume of
André's treatise, which he referred to in his presidential address to
the Astronomical Society of France as an ``excellent book''
\citep{hp1902sa}. Similarly, in his review for the \emph{Bulletin
  astronomique}, Poincaré observed that this new volume ``cedes
nothing to the first'' with respect to the interest of its contents
\citep{hp1901bab}.

Here again, however, Poincaré corrected an error that involved, once
again, the principle of relative motion. In the case of a close
binary, André explained, when the companion eclipses the principal
star, its radial velocity vanishes, such that the observed radial
velocity of the principal fixes the velocity of the companion with
respect to the principal star \citep[79]{AndC1900}. Poincaré pointed out the obvious fact
that at the moment of conjunction the companion has null radial
velocity with respect to the principal. Consequently, observation of
spectral lines of the principal star at conjunction fixes only
the relative motion of the system's center of gravity with respect to
the Earth.

This was the only fault Poincaré found with André's second volume,
which he went on to present in detail. He reserved his most lyrical
prose for the section on close binaries, the method of observation of
which he singled out for praise. In fact, what he admired was just
André's dutiful summary of Myers' thesis. This part of Poincaré's
review merits a lengthy citation:
\begin{quote}
  We are pleased to see united and related all these recent discoveries,
  which over that last few years seem to give a glimpse into a
  whole new world. These new views, so endearing and seductive, remain
  in part hypothetical. There are, nonetheless, conclusions which at
  present appear to have acquired a high degree of probability. The
  variability of Algol-type stars appears to be due to eclipses, and that of
  $\beta$ Lyrae-type stars, to the considerable flattening of the two
  components which, during their rotation and orbital revolution,
  would present to the observer's eye at one moment a small section,
  and at another moment a large one. The former would be completed
  binary systems, the others, binary systems being
  formed.\footnote{``On aime à voir réunies et rapprochées toutes ces
    découvertes récentes qui depuis quelques années semblent nous
    ouvrir un aperçu sur un monde tout nouveau. Ces vues nouvelles, si
    attachantes et si séduisantes, sont encore en partie
    hypothétiques. La variabilité des étoiles du type d'Algol paraît
    due à des éclipses, celle des étoiles du type $\beta$ de la Lyre à
    l'aplatissement considérable des deux composantes qui, pendant
    leur rotation et leur révolution orbitale, présenteraient à
    l'oeil de l'observateur tantôt une section faible, tantôt une
    section considérable. Les premières seraient des systèmes binaires
    formés, les autres des systèmes binaires en voie de formation.''
    \citet[44]{hp1901bab}}
\end{quote}
The view expressed here by Poincaré, that classes of variables are
eclipsing binaries or proto-binaries, was shared by many theorists at
the time. Poincaré was likely aware of Vogel's admission, in his 1900
review of a decade of progress in stellar motion determination, that
Myers' orbit ``very satisfactorily represented'' the light curve of
$\beta$ Lyrae \citep{VogH1900}.\footnote{Similarly,
  \citet[112]{NewS1902a} adopted Myers' view of $\beta$ Lyrae and U
  Pegasi as eclipsing binaries, and added a third example: $\zeta$
  Herculis.}

Poincaré's public embrace of Myers' interpretation of the pear in his
glowing review of André's textbook undoubtedly
attracted notice among astronomers and astrophysicists. It may well
have come to the attention of Myers, who wrote to Poincaré in
September, 1901, some eight months after his review appeared in the
\emph{Bulletin}. In his letter, which is transcribed in the appendix,
Myers mentioned André's résumé of his research on $\beta$ Lyrae and U
Pegasi, and presented his plan to pursue this line of research, since
``by computing light curves based upon any of your other theoretical
forms -- perhaps the apiodal'' he might reproduce the light curves of
more variables. To further this project, Myers requested a copy of
Poincaré's publications on equilibrium forms of rotating fluid
masses. This would help him, Myers explained to Poincaré,
\begin{quote}
  to accomplish something which may be of interest to you as
  furnishing tangible proof of the existence in the universe of such
  equilibrium forms as your matchless pen has proved possible.
\end{quote}
Myers' request suggests that he was not yet fully acquainted with
Poincaré's scientific publications on equilibrium figures. This is
quite plausible, as his calculation of the light curve of $\beta$
Lyrae required only equilibrium figures of revolution, and not the
figures discovered by Jacobi or Poincaré.

We don't know if Poincaré responded to Myers' letter, or if he sent
the requested offprints. What we \emph{do} know, is that Myers
published no further on spectroscopic binaries, or on any other matter
of scientific research. From 1901 until his retirement in 1929, Myers
was professor of mathematics and astronomy in the College
of Education at the University of Chicago \citep[885]{Marq1943}, a
professional circumstance which would have limited his access to
research support and graduate students in astrophysics.

As for Poincaré, he was not done with the pear or his fission theory. His
lectures at the Sorbonne for the academic year 1900--1901 were devoted
to equilibrium figures of a rotating fluid mass, and were written up
for publication by
Léon Dreyfus \citep{hp1902fe}. Beginning in May,
1901, Darwin entertained correspondence with Poincaré on the topic of
exchange of stabilities, which lasted a full year, and led to several related
publications by both men.\footnote{See the annotated transcriptions of
  Darwin's and Poincaré's letters in \citet{hp2016cp}, and the
  introduction by Ralf Krömer to this fascinating exchange.}

\section{Conclusion}

When Poincaré took up the problem of determining the equilibrium
figures of a rotating fluid mass in the mid-1880s, the possibility of
observing a concrete realization of such figures in the solar system
or the stellar universe was distant, but perhaps not beyond
imagination. Spectroscopic binaries did not exist when Poincaré
published his first work on equilibrium figures, but within four
years, they did, and within ten years, Myers' model of $\beta$ Lyrae
led him -- and many other theorists -- to believe that it was not a
true binary star, but a proto-binary in the form of a Poincaré pear. Myers'
fellow doctoral student in Munich, and a rising star of German
astrophysics, Karl Schwarzschild was in full agreement, as he publicly
chided Poincaré for his initial reluctance to abandon the nebular
hypothesis. In France, soon after the discovery of spectroscopic binaries,
Poincaré brought his fission theory to the attention of a broad
readership, while Charles André wrote a textbook on Myers' genial
method of reducing light curves to orbital elements.

We have now seen what Poincaré's precise role was in the first fifteen
years or so of the ``intoxicating'' history of the pear in
cosmogony. No one would argue that either the Poincaré-Darwin fission
theory, or the pear itself stood as a necessary condition for Myers'
model of $\beta$ Lyrae. It also seems likely that André's presentation
of Myers' method of reducing $\beta$ Lyrae's light curve to orbital
elements was not wholly premised on Myers' computation of the system's
equipotential surfaces. Chandrasekhar, I believe, was right about
Poincaré's considerable power of persuasion. His somewhat negative
characterization of Poincaré's role in the history of the pear,
however, has obscured the importance of the Poincaré pear and of
Poincaré-Darwin fission theory
to astrophysical interpretation of a class of variable stars as
eclipsing binaries or proto-binaries, the prototype of which is the
pear-shaped $\beta$ Lyrae.

\section*{Annex: Letter from G. W. Myers to H. Poincaré}

\begin{flushright}
Sept.~24, 1901\\
Chicago Ills.~U.S.A. -- 6026 Monroe Ave.\\
\textsc{Chicago} University --- William R. Harper,
\textsc{President}\\
Professor of Astronomy and Mathematics --- School of Education
\end{flushright}

\noindent
Mr. H. Poincaré, Paris France

\smallskip
\noindent
My dear Sir:

\noindent
You have perhaps noticed in Professor André's book entitled
\emph{``Traité d'astronomie stellaire''} Vol II p 303 that my
discussions of $\beta$ Lyrae and U Pegasi both seem to point to a
concrete confirmation of your excellent work on rotating liquids.  I
am now curious to see if by computing light curves based upon any of
your other theoretical forms -- perhaps the apiodal -- I may be able
to represent the light curves of any other variables. To this end I
write to inquire whether you can put into my hands a copy of the paper
containing your discussion, or discussions, of your various forms of
rotating liquid masses in equilibrium. If you can do so, you will
greatly oblige me and help me, perhaps, to accomplish something which
may be of interest to you as furnishing tangible proof of the
existence in the universe of such equilibrium forms as your matchless
pen has proved possible.

\noindent
Most respectfully yours,

\noindent
G. W. Myers\footnote{Myers' one-page manuscript is in the Poincaré
  family archives, and may be consulted in digital form on the website
  \href{http://henripoincarepapers.univ-nantes.fr/corresphp/index.php?a=on&id=725&action=Chercher}{Henri Poincaré Papers}.}

\bigskip
\noindent
\emph{Acknowledgments}. This paper was presented at the Poincaré
meeting in Nancy organized by the Henri Poincaré Archives in July
2022. I thank the meeting participants, an anonymous referee, and the
editors for comments. I am also grateful to the Poincaré Estate for sharing
Henri Poincaré's correspondence and authorizing its publication.

\bibliography{mybib}
\end{document}